\documentclass[aps,prb,preprint,superscriptaddress,nobibnotes,amsmath,amssymb,amsfonts]{revtex4-1}

\usepackage{graphicx}
\usepackage{dcolumn}
\usepackage{bm}
\usepackage{multirow}
\usepackage{float}
\newcolumntype{C}[1]{>{\centering\arraybackslash}p{#1}}

\begin{document}

\title{Strain Engineering of Antimonene by a First-principles Study: \\Mechanical and Electronic Properties}

\author{Devesh R. Kripalani}
\affiliation{School of Mechanical and Aerospace Engineering, Nanyang Technological University, Singapore 639798, Singapore}
\affiliation{Infineon Technologies Asia Pacific Pte Ltd, Singapore 349282, Singapore}

\author{Andrey A. Kistanov}
\affiliation{School of Mechanical and Aerospace Engineering, Nanyang Technological University, Singapore 639798, Singapore}
\affiliation{Institute of High Performance Computing, Agency for Science, Technology and Research, Singapore 138632, Singapore}

\author{Yongqing Cai}
\email[]{caiy@ihpc.a-star.edu.sg}
\affiliation{Institute of High Performance Computing, Agency for Science, Technology and Research, Singapore 138632, Singapore}

\author{Ming Xue}
\affiliation{Infineon Technologies Asia Pacific Pte Ltd, Singapore 349282, Singapore}

\author{Kun Zhou}
\email[]{kzhou@ntu.edu.sg}
\affiliation{School of Mechanical and Aerospace Engineering, Nanyang Technological University, Singapore 639798, Singapore}


\begin{abstract}
Recent success in the experimental isolation and synthesis of highly stable atomically thin antimonene has triggered great interest into examining its potential role in nanoelectronic applications. In this work, we investigate the mechanical and electronic properties of monolayer antimonene in its most stable $\beta$-phase using first-principles calculations. The upper region of its valence band is found to solely consist of lone pair \textit{p}-orbital states, which are by nature more delocalized than the \textit{d}-orbital states in transition metal dichalcogenides, implying superior transport performance of antimonene. The Young's and shear moduli of $\beta$-antimonene are observed to be $\sim$25\% higher than those of bulk antimony, while the hexagonal lattice constant of the monolayer reduces significantly ($\sim$5\%) from that in bulk, indicative of strong inter-layer coupling. The ideal tensile test of $\beta$-antimonene under applied uniaxial strain highlights ideal strengths of 6 GPa and 8 GPa, corresponding to critical strains of 15\% and 17\% in the zigzag and armchair directions, respectively. During the deformation process, the structural integrity of the material is shown to be better preserved, albeit moderately, in the armchair direction. Interestingly, the application of uniaxial strain in the zigzag and armchair directions unveil direction-dependent trends in the electronic band structure. We find that the nature of the band gap remains insensitive to strain in the zigzag direction, while strain in the armchair direction activates an indirect-direct band gap transition at a critical strain of 4\%, owing to a band switching mechanism. The curvature of the conduction band minimum increases during the transition, which suggests a lighter effective mass of electrons in the direct-gap configuration than in the free-standing state of equilibrium. The work function of free-standing $\beta$-antimonene is 4.59 eV and it attains a maximum value of 5.07 eV under an applied biaxial strain of 4\%. The findings reported in this work provide fundamental insights into the mechanical behaviour and strain-tunable nature of the electronic properties of monolayer $\beta$-antimonene, in support of its promising role for future nanoelectromechanical systems and optoelectronic applications.
\end{abstract}


\maketitle

\section{Introduction}
Two-dimensional (2D) materials have garnered substantial scientific attention in recent years due to its promising role for future nanoelectromechanical systems (NEMS) and optoelectronic applications. Since the discovery of graphene in 2004, the search for other atomically thin candidates has motivated extensive works on diverse material systems across the periodic table. These include various Group IV elemental analogues like germanene\cite{CTsige09,LFsige11,ODge14} and silicene,\cite{DTsi05,CTsige09,LFsige11,VDsi12} hexagonal boron-nitride (h-BN),\cite{WThbn04,RFhbn10,SChbn10,KHhbn12} transition metal dichalcogenides (TMDs) such as molybdenum disulfide (MoS$_{2}$)\cite{ACmos11,AStmd12,CPmos12,DFtmd14} and novel van der Waals heterostructures.\cite{CYwfcn16} Lately, few-layer phosphorene has demonstrated great possibilities in NEMS and optoelectronics due to its robust, thickness-dependent finite direct band gap and eminent transport characteristics with high room temperature mobility.\cite{KCphos16,CKphos15,LYphos14,LNphos14,SEphos14,QXphos14} Owing to its puckered atomic structure and soft P-P bonds, a highly itinerant character of atomic vacancies has been proposed as a unique feature of phosphorene which may lead to superplastic deformation under moderate temperature.\cite{CKphos16}

More recently, another Group V elemental 2D material, known as antimonene (i.e. a single layer of antimony), has attracted significant academic interest due to its intriguing electronic and optical properties.\cite{PSgrpv17,APsbdft17,SLsbelec18} Furthermore, it has successfully been experimentally obtained via mechanical isolation from bulk antimony.\cite{AAsbexp16} High-quality few-layer antimonene nanosheets with large lateral dimensions have also been produced via liquid exfoliation\cite{GRsbexp16} and van der Waals epitaxy growth above various substrates such as germanium\cite{FWsbge17} and other 2D layered structures.\cite{JSsbexp16,WSsbexp17} Antimonene can exist in several possible phases ($\alpha$-, $\beta$-, $\gamma$-, $\delta$-), with the rhombohedral $\beta$-phase being its most stable, ground state configuration.\cite{AAsbexp16,AOsbdft15,WPsbdft15,ZYsbbands15} In this phase, antimonene adopts a puckered honeycomb lattice, bearing close resemblance to silicene. However, as opposed to the semimetal character of bulk antimony, antimonene possesses an indirect band gap. Experiments have also established that antimonene remains highly stable in ambient conditions, showing little or no signs of degradation over periods of months.\cite{AAsbexp16} This is in stark contrast to the poor stability of its predecessor - phosphorene. While great strides have been made towards understanding the optical characteristics,\cite{AZsbopt17} doping possibilities,\cite{UAsbdope16,MCsbdope17,KCsbdope18} functionalization-induced charge dynamics,\cite{AAsbfunc17,ZZsbox17} many-body physics\cite{WHsbmb17} and quantum-confinement effects\cite{NCsbrbt17} of antimonene, there remain lots of mystery surrounding this new 2D material, especially with regard to its response to applied strains, temperature loads and external electric fields.

Particularly, in order to realize the exciting opportunities of ultrathin materials such as antimonene, it is imperative to understand how its electronic and optical properties can be modulated at the atomic level. A purely mechanical approach, known as strain engineering, which involves the deliberate application of controlled strains onto the nanostructure, presents itself as a useful technique for tailoring its optoelectronic properties.\cite{LYsteng14,PWsteng14,DHsteng14,JPsteng15} In practice, strains may be reproducibly introduced via the use of stretchable elastomeric substrates.\cite{KZsteng09,CAsteng13} Moreover, since nanostructures are generally known to withstand much higher applied strains than its bulk counterparts,\cite{CAsteng13,HDsteng15} strain engineering potentially serves as an effective method for improving the performance of 2D materials in a practical environment. While numerous theoretical works on antimonene have identified its band gap to be highly tunable under moderate biaxial tensile strain,\cite{PSgrpv17,WPsbdft15,ZYsbbands15,SLsbelec18} they do so within the traditional symmetry-preserving treatment of the material system. As such, the sensitivity of its electronic band structure to symmetry-breaking, direction-dominant strain is unknown and continues to remain an open question of interest for achieving new, non-conventional functionalities of antimonene. Additionally, the influence of strain on its work function has not been explored and may offer fresh ground for facilitating efficient charge injection and transport across antimonene-based heterojunctions and interfaces with metals.

In this work, via first-principles calculations, we comprehensively investigate the mechanical characteristics of monolayer $\beta$-antimonene ($\beta$-Sb) in the elastic regime and provide direct insight into the underlying physical mechanism during the deformation process under applied uniaxial tensile strain in the zigzag and armchair directions. The effect of uniaxial and biaxial tensile strain on its electronic band structure and work function is subsequently analysed to evaluate the potential of strain-engineered antimonene for future nanoelectronic applications.

\section{Computational Method}
In this work, spin-polarized first-principles calculations are performed within the framework of density functional theory\cite{KSexcorr65,PYdft94,SSdft09} (DFT) using the Vienna Ab initio Simulation Package\cite{KFvasp96} (VASP). The Perdew-Burke-Ernzerhof (PBE) exchange-correlation functional\cite{PBE96} is adopted under the generalized gradient approximation (GGA) and hybrid Heyd-Scuseria-Ernzerhof (HSE06) functional methods.\cite{HSE03,HSE06} Due to the layered structure of antimonene, corrections for van der Waals forces are accommodated using the DFT-D2 method of Grimme.\cite{GSd206} Spin-orbit coupling (SOC) is also considered in band structure calculations as its effect is known to be significant for heavy elements such as antimony, which lies in the 5\textsuperscript{th} row of the periodic table. A kinetic energy cutoff of 450 eV is selected for the plane wave basis set, while the reciprocal space is sampled with a 20 $\times$ 20 $\times$ 1 \textit{k}-point grid in the Brillouin zone using the Monkhorst-Pack method. The energy convergence criteria for electronic iterations is set at 10\textsuperscript{-6} eV and all structures are relaxed until the maximum Hellmann-Feynman force per atom is smaller than 0.01 eV/\AA. Periodic boundary conditions are applied in the zigzag (\textit{x}) and armchair (\textit{y}) in-plane directions, whereas free boundary conditions are enforced in the normal (\textit{z}) direction by introducing a vacuum separation distance of 20 \AA\ to eliminate spurious interactions between replicate slabs.

\begin{figure}
\includegraphics[width=0.45\columnwidth]{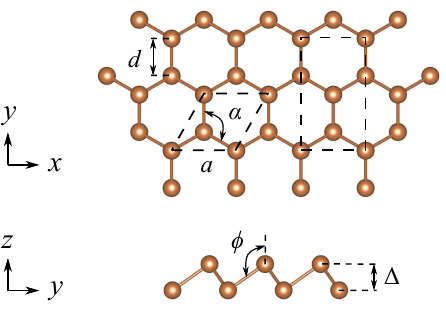}
\caption{The structural configuration of monolayer antimonene in its $\beta$-phase. \label{monoSb}}
\end{figure}

A single layer of antimonene in its $\beta$-phase is predicted to exhibit a stable, buckled honeycomb structure,\cite{AAsbexp16,AOsbdft15,WPsbdft15,ZYsbbands15} as shown in Fig. \ref{monoSb}. We report a hexagonal lattice constant \textit{a} of 4.07 \AA\ and a buckling height $\Delta$ of 1.66 \AA\ after full structural optimization. The cohesive energy per atom \textit{E}\textsubscript{c} is calculated to be 2.835 eV/atom, while the Sb-Sb bond length \textit{d} and inter-bond dihedral angle $\alpha$ are obtained as 2.88 \AA\ and 89.98$^\circ$, respectively. We also denote the angle formed between the Sb-Sb bond and the normal direction as $\phi$, which is found to be 125.28$^\circ$ for the relaxed monolayer. The calculated parameters in this work show good agreement with that from other theoretical publications\cite{WPsbdft15,AOsbdft15} and are summarized in Table \ref{relaxed_params}.

Throughout this work, strained $\beta$-Sb monolayers are simulated using an orthorhombic supercell approach, with the zigzag (armchair) direction aligned along the \textit{x} (\textit{y}) axis (see Fig. \ref{monoSb}). We denote the axial strain (stress) in the \textit{i} direction as $\epsilon_{i}$ ($\sigma_{i}$), and the shear strain (stress) in the \textit{ij}-plane as $\gamma_{ij}$ ($\tau_{ij}$). By convention, positive values of axial strain represent tension, while negative values refer to compression. In the case of applying uniaxial strain, the lattice constant in the transverse in-plane direction is fully relaxed in order to minimize the forces acting in this direction.

\begin{figure*}
\includegraphics[width=0.95\textwidth]{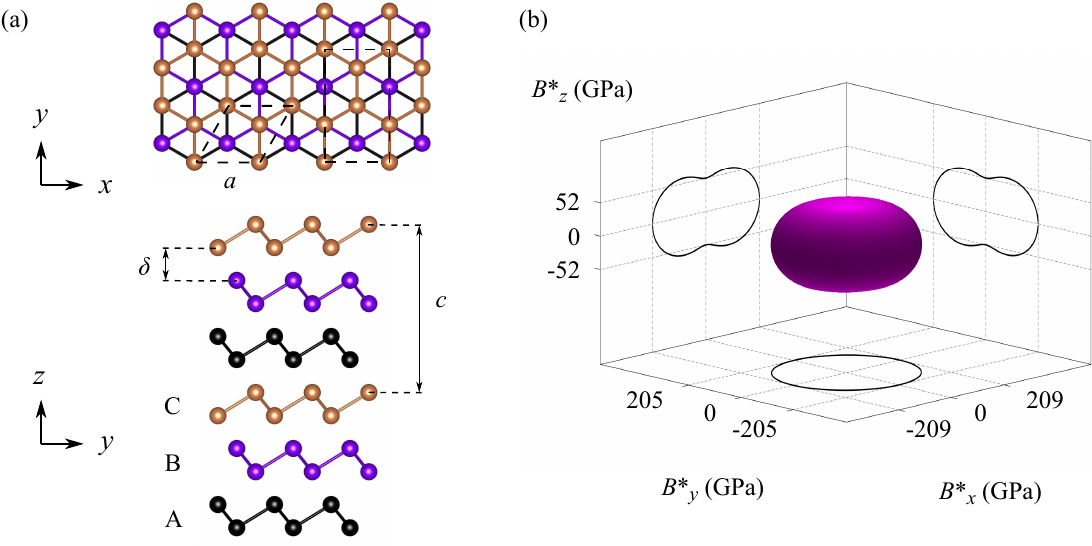}
\caption{(a) The structural configuration of bulk antimony and (b) its directional bulk modulus under hydrostatic pressure, represented on the orthogonal axes \textit{B}$_{i}^*$, where \textit{B}$_{i}^*$ = \textit{B}$_{\textbf{u}}$\textit{l}$_i$. The contour lines projected on the translated \textit{B}$_{i}^*$\textit{B}$_{j}^*$-planes illustrate the directional bulk modulus for the special cases of \textit{B}$_{k}^*$ = 0 (i.e. \textit{l}$_k$ = 0). \label{bulkSb}}
\end{figure*}

\section{Results and Discussion}
\subsection{Mechanical properties of bulk antimony}
Before we examine antimonene monolayers, the structure and mechanical properties of bulk antimony are first explored. In its bulk form, antimony adopts a hexagonal crystal lattice with space group \textit{R}$\bar{3}$\textit{m} (No. 166), as shown in Fig. \ref{bulkSb}a. Each unit cell consists of three $\beta$-Sb monolayers arranged in its ABC-stacked configuration. The three nearest neighbours of each Sb atom form intra-layer Sb-Sb bonds with lengths 2.92 \AA,\ while the three second-nearest neighbours form inter-layer Sb-Sb bonds with lengths 3.30 \AA.\ We report lattice constants \textit{a} and \textit{c} of 4.31 \AA\ and 11.09 \AA,\ respectively. The cohesive energy per atom \textit{E}\textsubscript{c} is calculated to be 3.117 eV/atom, while the inter-layer separation $\delta$ is obtained as 2.16 \AA.\ These calculated parameters are noted to be consistent with that from other experimental\cite{BCsb63,CKssp04} and theoretical\cite{AOsbdft15} works, and are summarized in Table \ref{relaxed_params}.

\begingroup
\squeezetable
\begin{table*}[t!]
\caption{ The relaxed geometric parameters, elastic constants and cohesive energy of bulk antimony and monolayer $\beta$-antimonene, as calculated in this work. \label{relaxed_params}}
\begin{ruledtabular}
{\renewcommand{\arraystretch}{1.5}
\begin{tabular}{cccccccc}
\multirow{2}{*}{System} & \multicolumn{2}{c}{\multirow{2}{*}{Geometric Parameters}} & \multicolumn{4}{c}{Elastic Constants} & {\textit{E}\textsubscript{c}} \\ \cline{4-7}
& & & \multicolumn{2}{c}{In-plane (\textit{xy})} & \multicolumn{2}{c}{Normal (\textit{z})} & (per atom) \\ \hline
Bulk & \textit{a} = 4.31 \AA & \textit{c} = 11.09 \AA & \textit{E}$_{x}$ = 83 GPa & \textit{E}$_{y}$ = 79 GPa & \textit{E}$_{z}$ = 33 GPa & & 3.117 eV \\
antimony & $\delta$ = 2.16 \AA & & \textit{G}$_{xy}$ = 35 GPa & & \textit{G}$_{xz}$ = 37 GPa & \textit{G}$_{yz}$ = 37 GPa & \\
& & & $\nu_{xy}$ = 0.16 & $\nu_{yx}$ = 0.15 & $\nu_{zx}$ = 0.18 & $\nu_{zy}$ = 0.19 & \\
& & & $\nu_{xz}$ = 0.44 & $\nu_{yz}$ = 0.46 & & & \\
& & & & & & & \\
Monolayer & \textit{a} = 4.07 \AA & & \textit{E}$_{x}$ = 104 GPa & \textit{E}$_{y}$ = 104 GPa & & & 2.835 eV \\
$\beta$-Sb & $\Delta$ = 1.66 \AA & \textit{d} = 2.88 \AA & \textit{G}$_{xy}$ = 43 GPa & & & & \\
& $\alpha$ = 89.98$^\circ$ & $\phi$ = 125.28$^\circ$ & $\nu_{xy}$ = 0.22 & $\nu_{yx}$ = 0.22 & & & \\
& & & $\nu_{xz}$ = 0.35 & $\nu_{yz}$ = 0.35 & & &
\end{tabular}}
\end{ruledtabular}
\end{table*}
\endgroup

In the elastic regime, the mechanical properties of bulk antimony can be characterized based on its elastic constants. Elastic constants are fundamental mechanical properties of crystalline materials and are valid within the limits for which Hooke's Law is obeyed. They describe the deformation response of a particular material under an externally applied state of stress. The generalized stress-strain relation for a three-dimensional crystalline solid is given by Eqn. \ref{ss3D} in the form of $\bm{\epsilon}$ = \textbf{S}$\bm{\sigma}$. Alternatively, this equation may be rewritten as $\bm{\sigma}$ = \textbf{C}$\bm{\epsilon}$, where \textbf{C} = \textbf{S}$^{-1}$. Here, \textbf{S} and \textbf{C} represent the compliance and stiffness matrices, respectively. They are 6 $\times$ 6 symmetric and contain the material-dependent elastic constants. In this work, \textit{S}$_{pq}$ (\textit{C}$_{pq}$) is used to concisely denote the individual elements of the compliance (stiffness) matrix, while \textit{E}$_{i}$ is the Young's modulus in the \textit{i} direction, $\nu_{ij}$ is the Poisson's ratio due to an applied (response) strain in the \textit{i} (\textit{j}) direction and \textit{G}$_{ij}$ is the shear modulus in the \textit{ij}-plane.
{\small
\begin{align}\label{ss3D}
& \begin{bmatrix}
\epsilon_{x} & \epsilon_{y} & \epsilon_{z} & \gamma_{xy} & \gamma_{xz} & \gamma_{yz}
\end{bmatrix}^\textbf{T}
 = \textbf{S}
\begin{bmatrix}
\sigma_{x} & \sigma_{y} & \sigma_{z} & \tau_{xy} & \tau_{xz} & \tau_{yz}
\end{bmatrix}^\textbf{T} \nonumber \\
& \text{where} \\
& \textbf{S} =
\begin{bmatrix}
{1}/{\textit{E}_{x}} & -{\nu_{yx}}/{\textit{E}_{y}} & -{\nu_{zx}}/{\textit{E}_{z}} & 0 & 0 & 0 \\
-{\nu_{xy}}/{\textit{E}_{x}} & {1}/{\textit{E}_{y}} & -{\nu_{zy}}/{\textit{E}_{z}} & 0 & 0 & 0 \\
-{\nu_{xz}}/{\textit{E}_{x}} & -{\nu_{yz}}/{\textit{E}_{y}} & {1}/{\textit{E}_{z}} & 0 & 0 & 0 \\
0 & 0 & 0 & {1}/{\textit{G}_{xy}} & 0 & 0 \\
0 & 0 & 0 & 0 & {1}/{\textit{G}_{xz}} & 0 \\
0 & 0 & 0 & 0 & 0 & {1}/{\textit{G}_{yz}}
\end{bmatrix} \nonumber
\end{align}
}

In order to evaluate the elastic constants of bulk antimony, we first sample the energy surface of the material under various strain configurations across the range -1.5\% $\leq$ $\epsilon_{i}$ $\leq$ 1.5\% and -3\% $\leq$ $\gamma_{ij}$ $\leq$ 3\%. For small deformations, the strain energy \textit{E}\textsubscript{s} exhibits a quadratic dependence on the applied strains, given by Eqn. \ref{Es_3D}. \textit{E}($\epsilon$) and \textit{E}$_{0}$ refer to the total energy of the strained and relaxed system respectively, while \textit{V}$_{0}$ is the equilibrium volume of the simulation cell.
{\small
\begin{align}\label{Es_3D}
\textit{E}\textsubscript{s} = \textit{E}(\epsilon)-\textit{E}_{0} &= \frac{V_0}{2}\bigg[C_{11}\epsilon_{x}^2 + C_{22}\epsilon_{y}^2 + C_{33}\epsilon_{z}^2 \nonumber \\
& \quad + 2C_{12}\epsilon_{x}\epsilon_{y} + 2C_{23}\epsilon_{y}\epsilon_{z} + 2C_{13}\epsilon_{x}\epsilon_{z} \nonumber \\
& \quad + C_{44}\gamma_{xy}^2 + C_{55}\gamma_{xz}^2 + C_{66}\gamma_{yz}^2\bigg]
\end{align}
}

By performing a fit to the strain energy surface, we obtain the stiffness matrix \textbf{C} for bulk antimony, as shown in (\ref{C6}). We then determine its elastic constants, which are presented in Table \ref{relaxed_params}. 
{\small
\begin{gather}\label{C6}
\textbf{C} =
\begin{bmatrix}
96.46 & 24.90 & 21.77 & 0 & 0 & 0 \\
 & 93.34 & 22.19 & 0 & 0 & 0 \\
 & & 41.02 & 0 & 0 & 0 \\
 & & & 34.86 & 0 & 0 \\
\multicolumn{3}{c}{Sym.} & & 36.66 & 0 \\
 & & & & & 36.96
\end{bmatrix}
\end{gather}
}

Our results reveal that bulk antimony behaves as a transversely isotropic material in the \textit{xy}-plane, with a substantially lower Young's modulus in the \textit{z} direction (\textit{E}$_{x}$ $\simeq$ \textit{E}$_{y}$ $\simeq$ 2.5\textit{E}$_{z}$). Its relatively high in-plane stiffness is a testament to the strong intra-layer bonding within the material. The values of the Poisson's ratio $\nu_{xz}$ and $\nu_{yz}$ are found to be large, which is a general indication that in-plane stress can trigger significant normal deformation. It is also of prime interest to investigate the compressibility of bulk antimony under hydrostatic pressure, given by the relations in (\ref{KBbulk}) at the same level of theory. The bulk modulus \textit{K} is computed to be 35 GPa, while the directional bulk modulus \textit{B}$_\textbf{u}$ is represented on Fig. \ref{bulkSb}b. Here, \textit{B}$_\textbf{u}$ is a measure of linear compressibility in an arbitrary direction and is defined by the unit vector \textbf{u} = \textit{l}$_{1}$\textit{\textbf{\^i}}+\textit{l}$_{2}$\textit{\textbf{\^j}}+\textit{l}$_{3}$\textit{\textbf{\^k}}, where \textit{l}$_{i}$ are the direction cosines.
{\small
\begin{align}\label{KBbulk}
K = \bigg[\sum_{p=1}^{3} \sum_{q=1}^{3} S_{pq}\bigg]^{-1} ,\quad
B_\textbf{u} = \bigg[\sum_{p=1}^{3} \bigg(\sum_{q=1}^{3} S_{pq}\bigg)l_p^2\bigg]^{-1}
\end{align}
}

The bulk elastic properties derived in this work are congruous with those readily available as part of the Materials Project.\cite{JCmatproj15} The predicted Young's, shear and bulk moduli also agree well with experimentally observed values for polycrystalline samples of bulk antimony,\cite{Hmech56} which have been reported as $\sim$55 GPa, $\sim$20 GPa and $\sim$36 GPa, respectively. The non-spherical profile of the directional bulk modulus (Fig. \ref{bulkSb}b) is a reflection of considerable anisotropy in out-of-plane directions (i.e. \textit{l}$_3$ $\neq$ 0). In its reference orientation, we note that \textit{B$_{x}$} and \textit{B$_{y}$} are 209 GPa and 205 GPa respectively, while \textit{B$_{z}$} has a value of 52 GPa, which is $\sim$75\% smaller than the in-plane values. This highlights the relative ease of deformation in the normal direction due to the presence of weak inter-layer bonding within the material. In all, the analysis provided with this section presents useful theoretical guidelines for practical applications involving the mechanical isolation of antimonene from polycrystalline aggregates of bulk antimony.

\subsection{Characterization of $\beta$-Sb in the elastic regime}
Monolayer $\beta$-Sb presents itself as a 2D system and may be considered to be in a state of plane stress (i.e. $\sigma_{z}$ = $\tau_{xz}$ = $\tau_{yz}$ = 0). Hence, its stress-strain and \textit{E}\textsubscript{s}-strain relations take on simplified forms, given by Eqn. \ref{ss2D} and Eqn. \ref{Es_2D}, respectively. In Eqn. \ref{Es_2D}, \textit{A}$_{0}$ is the equilibrium in-plane area of the simulation cell, while \textit{t}$_{0}$ refers to the thickness of monolayer $\beta$-Sb in its relaxed configuration.
{\small
\begin{align}\label{ss2D}
\begin{bmatrix}
\epsilon_{x} \\ \epsilon_{y} \\ \gamma_{xy}
\end{bmatrix}
= \textbf{S}
& \begin{bmatrix}
\sigma_{x} \\ \sigma_{y} \\ \tau_{xy}
\end{bmatrix}, \quad \epsilon_{z} = -\frac{\nu_{xz}}{\textit{E}_{x}}\sigma_{x}-\frac{\nu_{yz}}{\textit{E}_{y}}\sigma_{y} \\
\text{where} \ \textbf{S} =
& \begin{bmatrix}
{1}/{\textit{E}_{x}} & -{\nu_{yx}}/{\textit{E}_{y}} & 0 \\
-{\nu_{xy}}/{\textit{E}_{x}} & {1}/{\textit{E}_{y}} & 0 \\
0 & 0 & {1}/{\textit{G}_{xy}} \\
\end{bmatrix} \nonumber
\end{align}
}
{\small
\begin{align}\label{Es_2D}
\textit{E}\textsubscript{s} = \textit{E}(\epsilon)-\textit{E}_{0} =  \frac{A_0t_0}{2}\bigg[C_{11}\epsilon_{x}^2 + C_{22}\epsilon_{y}^2 + 2C_{12}\epsilon_{x}\epsilon_{y} + C_{44}\gamma_{xy}^2\bigg]
\end{align}
}

Following the same procedure from the previous section, we obtain the stiffness matrix \textbf{C} for monolayer $\beta$-Sb, as shown in (\ref{C3}). The calculated elastic constants for this material are listed in Table \ref{relaxed_params}. With reference to its bulk form, it is evident that monolayer $\beta$-Sb continues to retain its isotropic nature in the elastic regime, while the strength of its intra-layer bonds is enhanced. Quantitatively, we find that the Young's and shear moduli of the monolayer are $\sim$25\% higher than those of bulk antimony.
{\small
\begin{gather}\label{C3}
\textbf{C} =
\begin{bmatrix}
109.49 & 23.66 & 0 \\
 & 109.42 & 0 \\
Sym. & & 42.71
\end{bmatrix}
\end{gather}
}

Moreover, we observe that monolayer $\beta$-Sb has a relatively low Young's modulus as compared to other atomically thin materials such as graphene (1.0 TPa)\cite{LWgraphene08} and MoS$_{2}$ (0.33 TPa).\cite{CPmos12} Its low elastic rigidity is comparable to h-BN (0.25 TPa)\cite{SChbn10} and phosphorene (zigzag: 0.166 TPa, armchair: 0.044 TPa).\cite{QXphos14} This may be attributed to antimony being in the 5\textsuperscript{th} row of the periodic table and thereby, forming longer, yet weaker bonds than its 2\textsuperscript{nd} row counterparts. Hence, with its superior flexibility, $\beta$-Sb appears to be a promising candidate for practical strain engineering applications.

\subsection{Ideal tensile deformation of $\beta$-Sb}
The application of controlled strain forms the cornerstone of modern nanoscale device engineering as an effective means of tailoring the optical and electronic properties of materials at the atomic level. In this section, we explore the mechanical deformation of monolayer $\beta$-Sb under applied uniaxial tensile strain in the zigzag and armchair directions independently.

\begin{figure*}
\includegraphics[width=0.95\textwidth]{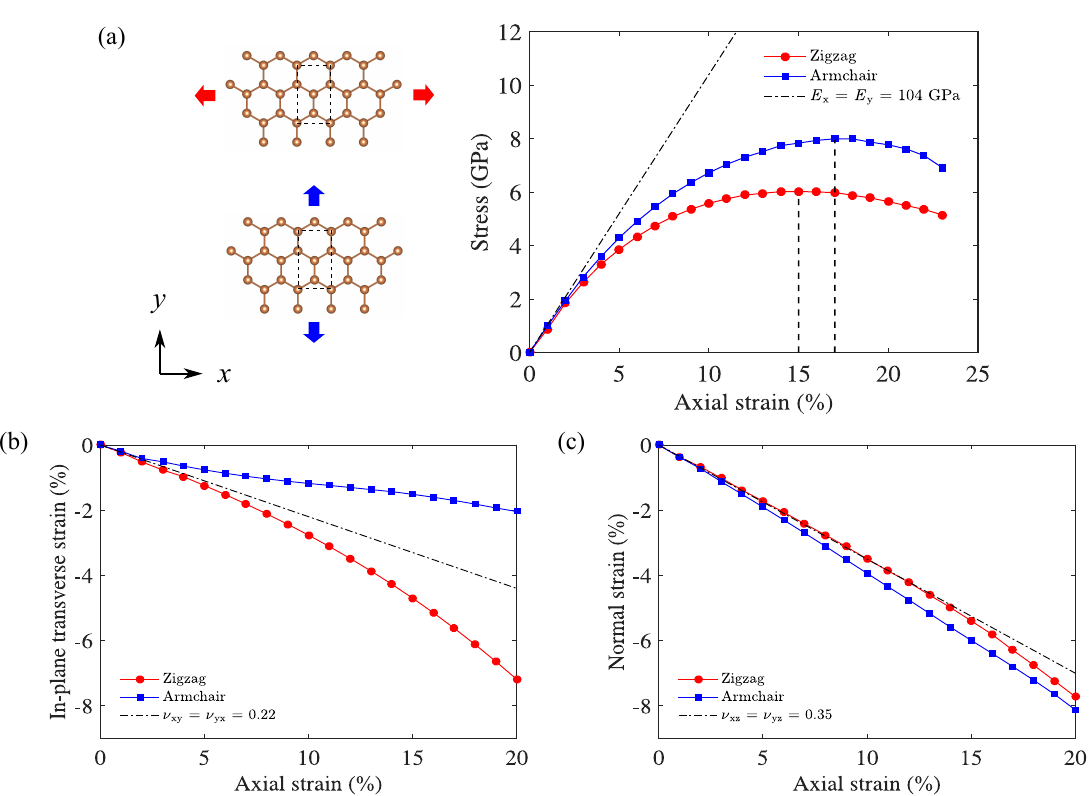}
\caption{(a) A schematic atomic model of monolayer $\beta$-antimonene being subjected to in-plane uniaxial strain in the zigzag and armchair directions (left panel). The stress-strain relationship (right panel), (b) in-plane transverse strain response and (c) normal strain response for monolayer $\beta$-antimonene under applied uniaxial tensile strain in the zigzag (red line) and armchair (blue line) directions. The dash-dotted lines indicate the fitted values of the elastic constants (see Table \ref{relaxed_params}). \label{stress_strain}}
\end{figure*}

Since the strain energy refers to the work done on the material by an externally applied stress, we can derive an expression for the engineering stress $\sigma_{i}$* as a function of uniaxial strain $\epsilon_{i}$ in the \textit{i} direction, given by Eqn. \ref{engstress}. Figure \ref{stress_strain} depicts the predicted stress-strain relationship and transverse relaxation response of monolayer $\beta$-Sb for a strain range of practical interest.
{\small
\begin{align}\label{engstress}
\sigma_{i}\text{*}(\epsilon_{i}) &= \frac{1}{\textit{A}_{0}t_{0}} \bigg[ \frac{1}{1+\epsilon_{i}} \bigg] \bigg[ \textit{E}\textsubscript{s}'(\epsilon_{i}) - \frac{\textit{E}\textsubscript{s}(\epsilon_{i})}{1+\epsilon_{i}} \bigg]
\end{align}
}

Monolayer $\beta$-Sb deforms approximately linearly and isotropically up to a strain of 2\%, beyond which anisotropy sets in. Outside the elastic regime, we find that the material demonstrates moderately greater resistance to deformation in the armchair direction than in the zigzag direction. The ideal strengths of $\beta$-Sb are 6 GPa and 8 GPa, corresponding to critical strains of 15\% and 17\% in the zigzag and armchair directions, respectively. The theoretical stress-strain behaviour of this material has also been reported by Wang et al.\cite{WPsbdft15} using the procedure of Wei and Peng,\cite{QXphos14} in which higher ideal strengths of 10 GPa and 11 GPa were identified at similar critical strains of 15\% and 18\% in the zigzag and armchair directions, respectively. Comparatively, we find that the proposed formulation adopted in this work (Eqn. \ref{engstress}) reproduces the elastic regime and structural anisotropy of $\beta$-Sb in a more consistent and accurate manner. The use of this formulation is also widely applicable and can be seamlessly extended to the study of other material systems of interest.

Moreover, as shown in Fig. \ref{stress_strain}b, the application of uniaxial tensile strain in the zigzag (armchair) direction will lead to a corresponding contraction in its transverse armchair (zigzag) direction, implying a positive Poisson's ratio (Table \ref{relaxed_params}). While the degree of transverse lattice contraction is directionally independent for low to moderate strain, its trend varies between the zigzag and armchair directions at higher strain levels, in which we observe that relaxation occurs more freely in the zigzag direction (see dashed line in Fig. \ref{stress_strain}b). Applied in-plane uniaxial tensile strain also induces out-of-plane compression of the lattice buckling height in the normal direction, as indicated in Fig. \ref{stress_strain}c.

\begin{figure*}
\includegraphics[width=0.95\textwidth]{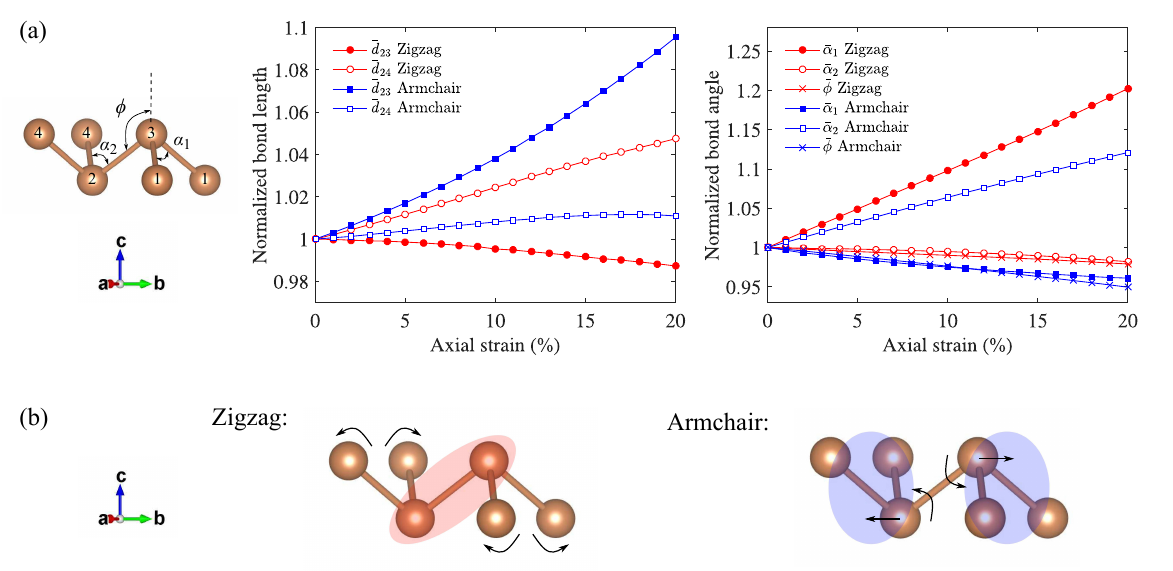}
\caption{(a) The variation of the structural geometry of monolayer $\beta$-antimonene under applied uniaxial tensile strain in the zigzag (red line) and armchair (blue line) directions. The bond lengths and angles are normalized with respect to their respective equilibrium values calculated in Table \ref{relaxed_params}. (b) A schematic view of the rigid zones (shaded) and dominant structural response (arrows) during uniaxial tensile deformation in the zigzag and armchair directions. Note: (\textbf{\emph{abc}}) $\equiv$ (\textbf{\emph{xyz}}). \label{BLA}}
\end{figure*}

In order to gain further insight into the energetics of $\beta$-Sb under applied uniaxial strain, we examine the geometric characteristics of the orthorhombic cell during structural deformation. Changes in the layer's bonding configuration are defined by several key parameters, as shown in Fig. \ref{BLA}a. The variation in Sb-Sb bond lengths (\textit{d}$_{23}$ and \textit{d}$_{24}$), dihedral angles ($\alpha_{1}$ and $\alpha_{2}$) and angle $\phi$ are plotted in Fig. \ref{BLA}a, providing a new perspective of the geometric evolution during uniaxial tensile deformation. It is to be noted that \textit{d}$_{31}$ = \textit{d}$_{24}$ via geometric symmetry.

In the zigzag direction, \textit{d}$_{23}$ and $\alpha_{2}$ undergo relatively small decrease, while \textit{d}$_{24}$ and $\alpha_{1}$ increase significantly. On the other hand, in the armchair direction, \textit{d}$_{23}$ and $\alpha_{2}$ experience dominant changes as compared to \textit{d}$_{24}$ and $\alpha_{1}$, which only show small fluctuations. In both directions however, $\phi$ does not appear to exhibit significant distortion. From this analysis, we are able to identify relatively rigid zones and its constituent bonds within the orthorhombic cell, as indicated by the shaded regions in Fig. \ref{BLA}b.

Despite having identical bond lengths (\textit{d}$_{23}$ = \textit{d}$_{24}$ = 2.88 \AA) and dihedral angles ($\alpha_{1}$ = $\alpha_{2}$ = 89.98$^\circ$) in its equilibrium configuration, applied uniaxial tensile strain in the zigzag and armchair directions preserve the rigidity of the Sb-Sb bonds differently. Hence, the onset of anisotropy beyond the elastic regime may be attributed to the extent of preserved Sb-Sb bond pairs, which is greater in the case of applied strain in the armchair direction. This leaves the armchair-strained configuration stronger than that of the zigzag-stained case, such that it yields at a slightly higher critical strain.

\begin{figure*}
\includegraphics[width=0.95\textwidth]{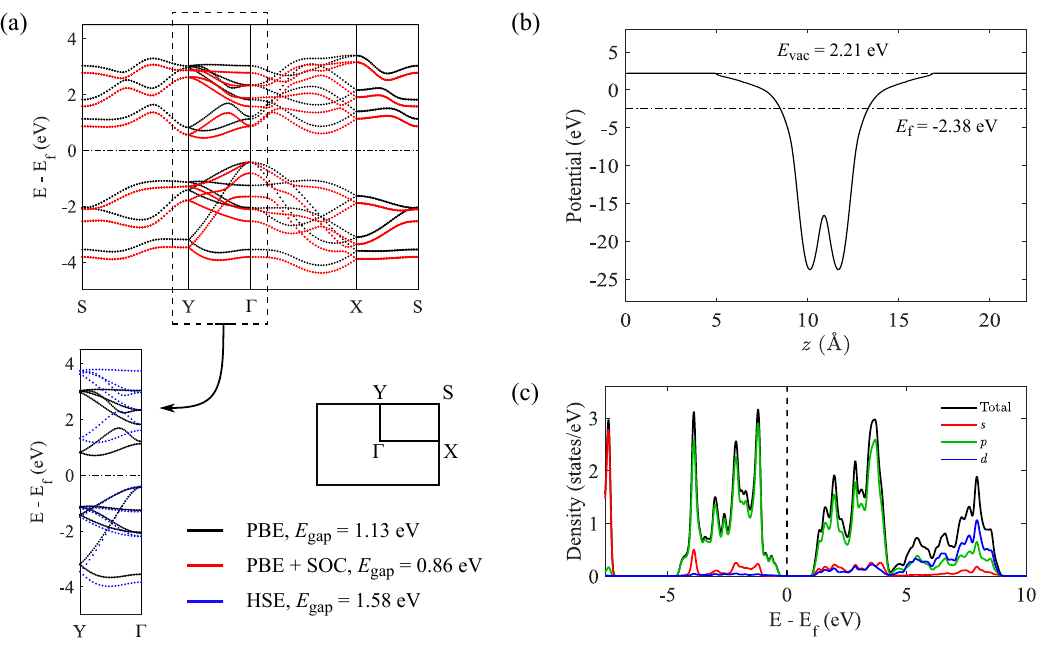}
\caption{(a) The electronic band structure of monolayer $\beta$-antimonene in its equilibrium configuration. The bands are calculated at the PBE level of theory (black lines) and corrected to include the effect of SOC (red lines). A separate comparison along the Y-$\Gamma$ path is also made using the HSE06 functional (blue lines) to obtain a more accurate estimate of the band gap \textit{E}\textsubscript{gap}. (b) The planar-averaged potential along the normal direction, where \textit{E}\textsubscript{f} and \textit{E}\textsubscript{vac} are the energy values at the Fermi and vacuum levels, respectively. (c) The total and orbital projected density of states of monolayer $\beta$-antimonene in its equilibrium configuration. The dashed line indicates the Fermi level. \label{elecprop_0}}
\end{figure*}

\subsection{Strain effect on the electronic properties of $\beta$-Sb}
In this section, we investigate the effect of both uniaxial and biaxial tensile strain on the electronic band structure and work function of monolayer $\beta$-Sb. The band structures are calculated along the high symmetry path S-Y-$\Gamma$-X-S for an orthorhombic cell at the PBE+SOC level of theory. In its equilibrium state, we find that the material is a semiconductor with an indirect band gap of 0.86 eV, which occurs between the conduction band minimum (CBM) at a point along the Y-$\Gamma$ direction and the valence band maximum (VBM) at the $\Gamma$ point. The effect of SOC on the band structure is found to be significant such that degeneracies at the top of the valence band are removed. As shown in Fig. \ref{elecprop_0}a (upper panel), the VBM splits by 0.40 eV and the predicted band gap reduces by $\sim$0.27 eV when SOC is considered.

Since the band gaps are known to be underestimated by standard PBE, in which lacks a complete description of the self-interaction of electrons, supplementary calculations are conducted using the HSE06 functional, which yields a wider indirect band gap of 1.58 eV, as shown in Fig. \ref{elecprop_0}a (lower panel). Nevertheless, due to the computationally demanding nature of such calculations, we limit the use of the hybrid functional for work function calculations. For the purpose of this study, we find that PBE+SOC calculations are sufficient for capturing the nature and critical transitions involving the band gap.

The work function is an important surface property, having strong implications on the band alignment and quality of electronic contact between esspecially dissimilar materials at an interface. It is defined as the minimum energy required to remove an electron from a solid to a point in vacuum immediately outside its surface. In this work, we determine the equilibrium work function of monolayer $\beta$-Sb to be 4.59 eV (Fig. \ref{elecprop_0}b). Also, we calculate the total and projected density of states, as shown in Fig. \ref{elecprop_0}c. In line with previously reported findings by Zhang et al.,\cite{ZYsbbands15} we see that the 5\textit{p} orbital states dominate the regions surrounding the CBM and VBM, while the contribution of \textit{s}-orbital states increase in the lower portion of the valence band. The upper region of the valence band ($\sim$0.5 eV below the VBM) primarily consist of lone pair \textit{p}-orbital states of antimony, which implies a high sensitivity of antimonene to external adsorbates. In comparison with the \textit{d}-orbital states of TMDs such as MoS$_{2}$, the sole presence of more highly delocalized \textit{p}-orbital states near the VBM also suggest high transport efficiency and excellent mobility in antimonene. In addition, the lower region of its conduction band features a hybridization of \textit{s}-, \textit{p}- and \textit{d}-orbital states, which favour light absorption with the coupling of excited \textit{p}-orbital states from the VBM.

\begin{figure*}
\includegraphics[width=1.00\textwidth]{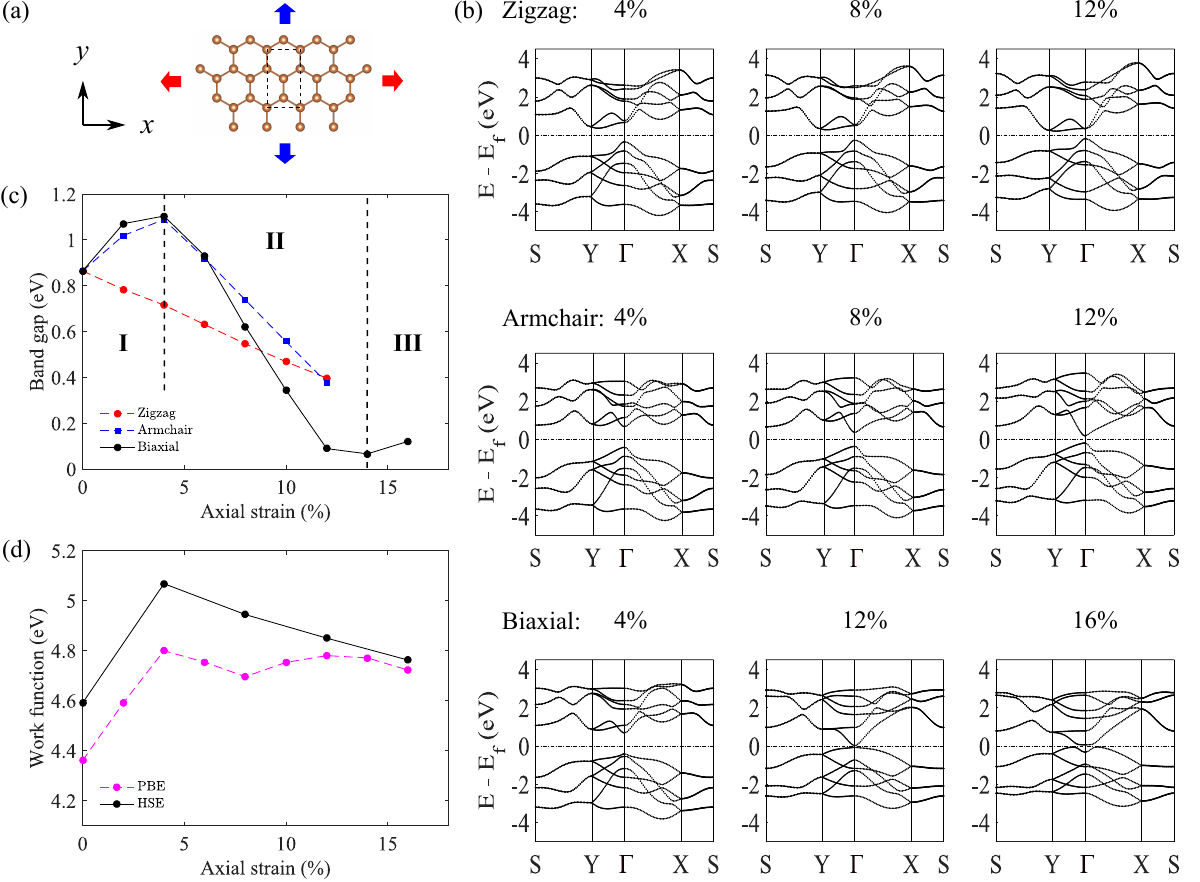}
\caption{(a) A schematic atomic model of monolayer $\beta$-antimonene being subjected to in-plane biaxial strain in both the zigzag and armchair directions. (b) The electronic band structure of monolayer $\beta$-antimonene under various applied uniaxial (zigzag and armchair) and biaxial strain configurations. (c) The variation of the electronic band gap under applied uniaxial and biaxial tensile strain. (d) The variation of the work function under applied biaxial tensile strain. \label{elecprop_strain}}
\end{figure*}

The strain-induced effects on the band structure of monolayer $\beta$-Sb are presented in Fig. \ref{elecprop_strain}b, while that on the band gap is captured in Fig. \ref{elecprop_strain}c. We find that applied tensile strain in the zigzag direction raises the VBM and lowers the CBM to an approximately equal extent with respect to the Fermi level. However, the indirect nature of the band gap remains well-preserved up to 12\% tension and is expected to prevail until the material structurally fails. During this process, the band gap of $\beta$-Sb, and thereby its semiconducting character, diminishes linearly with increasing strain in the zigzag direction.

In contrast, under applied tension in the armchair direction and likewise for the biaxial case, the situation is rather different. An indirect-direct band gap transition is predicted to occur at a moderate strain level of 4\%. Taking the biaxial case as a reference, we can identify three key regions of interest (I, II, III), as shown in Fig. \ref{elecprop_strain}c. In region I, below the critical level of 4\%, band-gap widening takes place, with the VBM remaining rather static with respect to the Fermi level. In region II, the band gap narrows monotonically until the CBM and VBM almost coincide at a second critical level of 13\%. Finally, in region III, the band gap experiences a direct-indirect transition and re-opens in a cusp-like manner. In both cases, the applied strain lifts the degeneracy at the $\Gamma$ point and triggers a downward shift of the second-lowest conduction band, while the VBM maintains its position at the $\Gamma$ point. The switching of energetic positions between the first- and second-lowest conduction bands at the $\Gamma$ point directly leads to the indirect-direct band gap transition. Notably, the curvature of the newly formed CBM at the $\Gamma$ point is much larger than that of free-standing antimonene, implying that the indirect-direct transition is accompanied by a dramatic drop in the effective mass of electrons, which enhances the transport efficiency of antimonene. While the biaxial-strain-driven indirect-direct transition has formerly been predicted in other works,\cite{PSgrpv17,WPsbdft15,ZYsbbands15} in our study, we essentially show that the application of armchair-dominant strain is both a necessary and sufficient condition for inciting the direct band gap in antimonene.

We also examine the variation of the work function of monolayer $\beta$-Sb for the biaxial case, as shown in Fig. \ref{elecprop_strain}d, with calculations based on the HSE functional giving more accurate results. The work function is observed to increase to a maximum value of 5.07 eV at 4\% strain and subsequently decrease under higher applied strain. Prior to the indirect-direct band gap transition at the critical value of 4\%, the work function increases due to reduced orbital splitting of antimony. In this strain regime, the valence electrons are more tightly bound and localized within the material. However, at higher applied strain, structural instabilities are likely to dominate and the weakening of Sb-Sb bonds contribute towards the decrease in the work function. We thus establish that the work function of $\beta$-Sb is mechanically responsive by up to +10\% and may be strain-engineered to improve charge injection and transport efficiency across heterojunctions and interfaces with metals. This finding paves new routes for the integration of antimonene into next-generation NEMS and optoelectronic devices.

\section{Conclusion}
In this work, we have investigated the effects of tensile strain on the mechanical deformation and electronic properties of monolayer $\beta$-Sb. The considerable enhancement ($\sim$25\%) of its Young's and shear moduli, and significant shrinkage ($\sim$5\%) of its lattice relative to that of bulk antimony suggest the presence of strong inter-layer coupling. Moreover, our results indicate that $\beta$-Sb deforms isotropically in the elastic region up to 2\%, beyond which moderate anisotropy sets in. Geometric analysis unveil the direction-dependent rigid zones which influence the strength of the monolayer such that it demonstrates a slightly higher critical strain of 17\% in the armchair direction. Our study provides direct insight into the underlying physical mechanism during the deformation process, which is of key practical relevance for the design of antimonene-based nanoelectronic devices.

Furthermore, our calculations reveal that the upper region of its valence band solely consist of lone pair \textit{p}-orbital states, which are by nature more delocalized than the \textit{d}-orbital states in TMDs, implying superior transport performance of antimonene. We have also characterized the strain-tunable band structure and work function of monolayer $\beta$-Sb. While the nature of the band gap is insensitive to strain in the zigzag direction, strain in the armchair direction activates an indirect-direct band gap transition at a moderate strain level of 4\%. This important property (i.e. electronic anisotropy) caters to both experimental and theoretical interests as it suggests new, non-conventional functionalities of antimonene for future NEMS and optoelectronic applications. We attribute the indirect-direct transition to the exchange of energetic positions between the two lowest-lying conduction bands. It is accompanied by a pronounced increase in the curvature of the CBM, implying that strained antimonene exhibits especially light effective mass of electrons in its direct-gap configuration.

This work systematically addresses its main objective by providing a closer look at how the mechanical and electronic properties of antimonene monolayers can be tuned by both uniaxial and biaxial tensile strain. A well-documented understanding of the strain-induced effects in its fundamental orthogonal directions is necessary for supplementing available literature on atomically thin antimonene, while providing a theoretical basis of support for further experimental endeavours which seek to explore antimonene as a novel semiconductor material.

\begin{acknowledgments}
This work was supported by the Economic Development Board, Singapore and Infineon Technologies Asia Pacific Pte Ltd through the Industrial Postgraduate Programme with Nanyang Technological University, Singapore and the Ministry of Education, Singapore (Academic Research Fund TIER 1 - RG174/15). The computational calculations for this work was partially performed on resources of the National Supercomputing Centre, Singapore. A. A. Kistanov acknowledges the financial support from the Agency for Science, Technology and Research (A*STAR), Singapore.
\end{acknowledgments}

\bibliography{bib_manuscript}

\end{document}